\documentstyle[amssymb,aps,prl,twocolumn,epsfig]{revtex}
\begin{document}
\twocolumn[\hsize\textwidth\columnwidth\hsize\csname @twocolumnfalse\endcsname

\title{Spectral evolution in (Ca,Sr)RuO$_3$ near the Mott-Hubbard transition}
\author{J. S. Ahn,$^1$ J. Bak,$^1$ H. S. Choi,$^1$ T. W. Noh,$^{1,2}$ J. E. Han,$^3$
Yunkyu Bang,$^4$ J. H. Cho,$^5$ and Q. X. Xia$^6$}
\address{$^1$Department of Physics, Seoul National University, Seoul 151-742, Korea}
\address{$^2$LG Corporative Institute of Technology, Seoul 137-140, Korea}
\address{$^3$Max-Planck Institut f\"{u}r Festk\"{o}perforschung, Heisenbergstr. 1,
D-70569, Stuttgart, Germany}
\address{$^4$Department of Physics, Chonnam National University, Kwangju 500-757,
Korea}
\address{$^5$RCDAMP and Department of Physics, Pusan National University, Pusan
609-735, Korea}
\address{$^6$Superconductivity Technology Center, Los Alamos National Laboratory, Los
Alamos, NM 87545, USA}
\draft
\maketitle

\begin{abstract}
We investigated the optical properties of (Ca,Sr)RuO$_{3}$ films on the
borderline of a metal-insulator (M-I) transition. Our results show all of
the predicted characteristics for a metallic Mott-Hubbard system, including ({\it i}) a mass enhancement in dc-limit, ({\it ii}) an $U/2$\ excitation,
and ({\it iii}) an $U$\ excitation. Also, a self-consistency is found within
the Gutzwiller-Brinkman-Rice picture for the Mott transition. Our finding
displays that electron correlation should be important even in $4d$\
materials.
\end{abstract}


\vskip 0.5pc]

Correlation between electrons in transition and rare earth metal compounds
has attracted lots of attentions. 
In general, the more localized the electron wave function is, the stronger the
correlation effects are. As a result, correlation effects are believed to be
much more important in describing $3d$ \ electrons than $4d$\ or $5d$\
electrons.

A metal-insulator (M-I) transition driven by electron correlation was
proposed by Mott and subsequently investigated intensively.\cite
{Mott1990Edwards1995Husmann1996} Since the Hubbard model was proposed in
early 1960's, it has been widely accepted as the simplest model which can
describe correlation effects. Although the model is composed of only two
parameters, i.e. inter-site hopping energy $t$ ($=W/z$)\ and on-site Coulomb
repulsive energy $U$, it has not been exactly solved yet except for one
dimensional case. [$W$\ and $z$\ are the bandwidth and the coordination
number, respectively.] Up to several years ago, different approaches
provided limited insights into different aspects of the M-I transition.
However, recent theoretical progresses, including a slave-boson approach,
infinite dimension limit approaches with several techniques, and numerical
calculations for finite size systems, started to provide a coherent picture.%
\cite{Georges1996}

\begin{figure}[htbp]
\epsfig{file=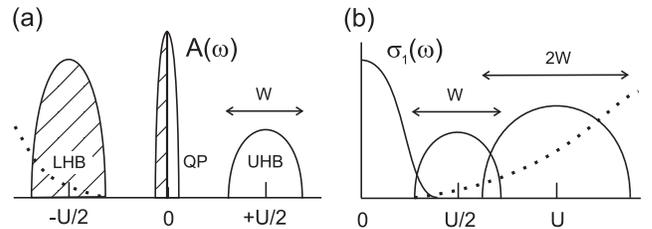,width=3.3in,clip=}
\vspace{2mm}
\caption{Schematic diagrams of (a) one particle spectral function and (b)
optical conductivity, for 4/6-filled metallic Mott-Hubbard system. Dotted
lines indicate the contributions from O($2p$) band.}
\label{schematic}
\end{figure}

According to the traditional Gutzwiller-Brinkman-Rice (GBR) picture,\cite
{Brinkman1970} the Mott M-I transition from a metallic side can be described
by narrowing and disappearing of a Fermi liquid quasi-particle (QP) band at
a critical value of correlation strength, $(U/W)_c$. Under this strong
renormalization, an effective mass, $m^{*}$,\cite{m*} of the QP is related
by: 
\begin{equation}
{\frac 1{{m^{*}}}}=1-{\frac{(U/W)^2}{{(U/W)_c^2}}}~.  \label{GBRmass}
\end{equation}
Recent theoretical works predict that one particle spectral function 
{\it A}($\omega $) for the metallic phase will be split into lower (LHB) and
upper (UHB) Hubbard bands, in addition to the QP band located at zero
frequency. Fig. \ref{schematic}(a) shows the schematic diagram of {\it A}($%
\omega $). Then, the corresponding optical conductivity spectra $\sigma
_1(\omega )$ can be easily predicted and displayed in Fig. \ref{schematic}%
(b). Note that $\sigma _1(\omega )$ in a metallic side has three
pronounced features: ({\it i}) a ``QP peak'' near zero frequency, ({\it ii}%
) an ``$U/2$ peak'' due to optical transitions between QP band and LHB (or
UHB), and ({\it iii}) an ``$U$ peak'' due to a transition between LHB and UHB.

In this letter, we will report optical properties of (Ca,Sr)RuO$_3$ films,
where four electrons occupy triply degenerate $t_{2g}$ levels. It is widely
accepted that some titanates and vanadates, which are usually approximated
as half-filled cases of a single orbital, can be described by the Hubbard
model. Most efforts to understand Mott-Hubbard physics have been focused on
such $3d$ oxides.\cite{Tokura1993,Makino1998} However, as far as we know,
such $3d$ oxides do not display all of the above-mentioned features for $%
1/m^{*}$ and $\sigma _1(\omega )$, which are expected for an ideal
Mott-Hubbard system. In ruthenates, there have been controversies between
band and correlation pictures.\cite
{Majin1997,Cao1997,Katsufuji1996,Kostic1998} Surprisingly enough, we found
that our (Ca,Sr)RuO$_3$ films, i.e. a 4/6-filled $4d$ electron system,
displays the mass enhancement and the optical features due to correlation
very clearly.

SrRuO$_3$ is known to be a bad metal which shows a ferromagnetic ordering at 
$T_c\sim $ 160 K. CaRuO$_3$ is also barely metallic, but it does not show
any magnetic ordering down to 4.2 K. Since CaRuO$_3$ has a narrower
bandwidth than SrRuO$_3$, it stands on the borderline of M-I transitions in
ternary Ru$^{4+}$ oxides.\cite{Cox1983} Therefore, if epitaxial CaRuO$_3$
thin films are fabricated, their electrical properties can be easily
controlled from metal to insulator through strain effect.\cite
{Rao1997Eom1992} Using this effect, we were able to prepare epitaxial
(Ca,Sr)RuO$_3$ thin films, where the Ca substitution makes them closer to the
M-I transition from metallic side.

\begin{figure}[htbp]
\epsfig{file=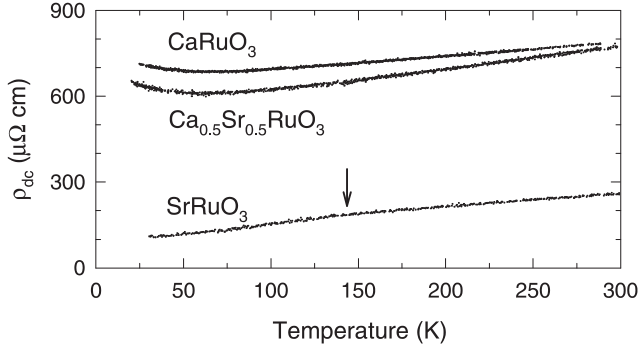,width=3.3in,clip=}
\vspace{2mm}
\caption{Resistivity versus temperature for (Ca,Sr)RuO$_3$. Arrow indicates
the Curie temperature of SrRuO$_3$. }
\label{resistivity}
\end{figure}

(Ca,Sr)RuO$_3$ films were grown on single crystal SrTiO$_3$(100) substrates
by pulsed laser deposition.\cite{Choi1998} Fig. \ref{resistivity} shows the
temperature dependent dc resistivity $\rho _{dc}$ curves. The SrRuO$_3$\
film shows a metallic behavior with a small slope change around 140 K, which
corresponds to its $T_c$ value. Both of the Ca$_{0.5}$Sr$_{0.5}$RuO$_3$ and
the CaRuO$_3$ films show barely metallic behaviors: their $\rho _{dc}$
values are close to about\ 1000 $\mu \Omega $ $cm$, which corresponds to the
Mott minimum metallic conductivity.\cite{Mott1990Edwards1995Husmann1996} The 
$\rho _{dc}$ value of the CaRuO$_3$ film is larger than its bulk value by a
factor of 3, due to the strain effect.

To obtain accurate optical properties of (Ca,Sr)RuO$_3$ films in a
wide frequency region of 0.23 $\sim $ 5.0 eV, we combined reflectance and
transmittance measurements with spectroscopic ellipsometry.\cite{Choi1998}
Between 0.23 and 3.0 eV, $\sigma _1(\omega )$ were obtained accurately from
reflectance and transmittance spectra using the Fresnel equations. Above 1.5
eV, $\sigma _1(\omega )$ were determined by spectroscopic ellipsometry.
In these methods, any extrapolation procedures, which are commonly used in
the Kramers-Kronig analysis of reflectance spectra,\cite{Kim1996} were not
required.

\begin{figure}[htbp]
\epsfig{file=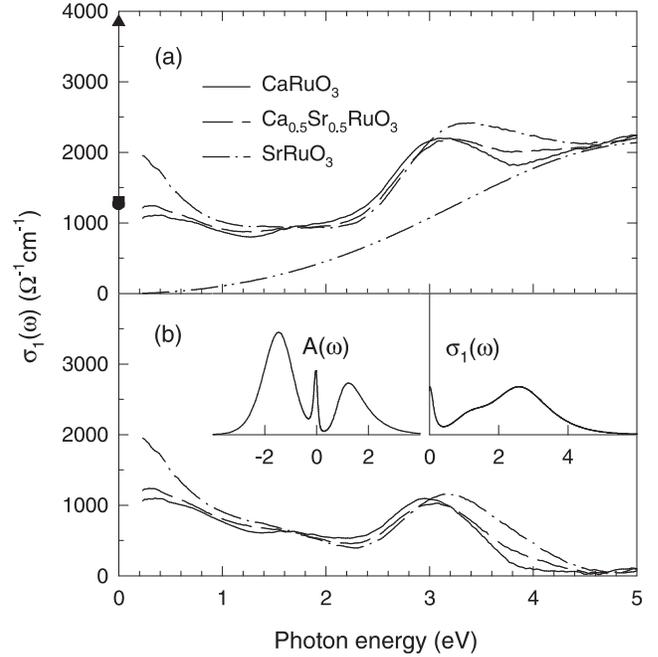,width=3.3in,clip=}
\vspace{2mm}
\caption{(a) Optical conductivity spectra of (Ca,Sr)RuO$_3$\ films at room
temperature. Symbols indicate dc values obtained from dc measurements. ($%
\blacktriangle $: SrRuO$_3$, $\blacksquare $: Ca$_{0.5}$Sr$_{0.5}$RuO$_3$,
and $\bullet $: CaRuO$_3$) Dashed-double-dotted line shows the charge
transfer transition between O($2p$)\ and Ru($4d$).\protect\cite{Lorentzian}
(b) Optical conductivity spectra after the charge transfer transition
contribution is subtracted. Inset: single particle spectral function and
optical conductivity calculated from a QMC simulation with $U/W$ $\sim $\
1.95 for the 4/6-filled triply degenerate orbitals.}
\label{cond}
\end{figure}

Figure \ref{cond}(a) shows $\sigma _1(\omega )$\ of the (Ca,Sr)RuO$_3$\
films. Below $\sim $ 1 eV, we can see the QP peaks whose dc limits agree
with the measured dc conductivity values, marked with symbols. As Ca
replaces Sr, the QP peak decreases. In the frequency region of 1 $\sim $ 2
eV, there are broad features of the $U/2$ peaks. Around 3 eV, the $U$ peaks
can be seen with a very broad background absorption.\cite{Fujioka1997} The
background comes from a charge transfer excitation between O($2p$) and Ru($%
4d $) with an energy higher than $U$,\cite{Lorentzian} and its estimated
contribution is indicated with the dashed--double-dotted line. Even with
this charge transfer contribution, this spectral feature resembles quite
closely to Fig. \ref{schematic}(b).

Figure \ref{cond}(b) shows $\sigma _1(\omega )$ after the background
absorption is subtracted. Note that the QP peaks are not changed very much
after the subtraction. From these spectra, $U$\ and $W$\ are experimentally
determined from the center and the half-width of the $U$\ peak: values of \{$%
U$, $W$\}, in the unit of eV, are \{2.99 $\pm $\ 0.05, 1.16 $\pm $\ 0.07\}
for CaRuO$_3$, \{3.07 $\pm $\ 0.05, 1.20 $\pm $\ 0.04\} for Ca$_{0.5}$Sr$%
_{0.5}$RuO$_3$, and \{3.23 $\pm $\ 0.05, 1.31 $\pm $\ 0.04\} for SrRuO$_3$.
As Sr replaces Ca, $U$ and $W$ increase systematically, where the relative
change in $W$ is larger than that of $U$. Therefore, {\it the (Ca,Sr)RuO}$_3$
{\it \ films can be considered effectively as a bandwidth controlled
Mott-Hubbard system}.

\begin{figure}[htbp]
\epsfig{file=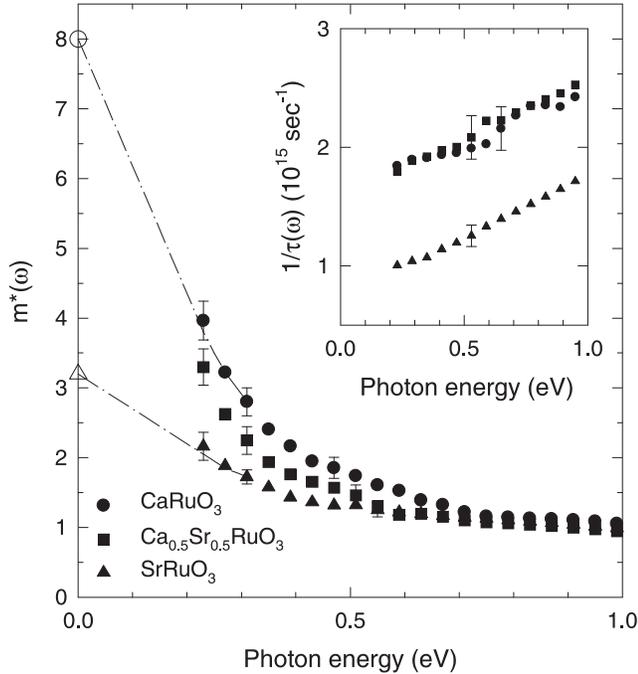,width=3.3in,clip=}
\vspace{2mm}
\caption{Low energy mass enhancement $m^{*}(\omega )$\ obtained from optical
data using the extended Drude model. $m^{*}$'s from specific heat measurements
are shown with $\bigcirc $\ (CaRuO$_3$)\protect\cite{Cox1983}\ and $%
\bigtriangleup $ ( SrRuO$_3$).\protect\cite{Cao1997} The dashed-dotted lines
are guidelines for eye. Inset shows corresponding frequency dependent
scattering rate, $1/\tau (\omega )$. }
\label{mass}
\end{figure}

The broad feature of the QP peaks could be described with the extended Drude
model,\cite{Webb1986} which provides informations on frequency dependent
mass $m^{*}(\omega )$ and scattering rate $1/\tau (\omega )$.\cite{Wp} As
shown in Fig. \ref{mass}, all of the samples showed mass enhancements as
frequency becomes lowered. Values of $m^{*}(\omega =0)$, i.e. $m^{*}$,
determined by specific heat measurements, for CaRuO$_3$\cite{Cao1997} and
SrRuO$_3$\cite{Cox1983} are marked with the open circle and triangle,
respectively. Since the carrier concentration is fixed for the (Ca,Sr)RuO$_3$
films, the enhancement of $m^{*}$ could be attributed to the correlation
effect. As Ca replaces Sr, $m^{*}$ becomes larger, indicating that CaRuO$_3$
is more close to the Mott M-I transition due to band narrowing. Similar
behaviors were observed for (Sr,La)TiO$_3$, which is a band filling
controlled Mott-Hubbard system.\cite{Tokura1993} On the other hand, such a
mass enhancement was not optically observed for (Ca,Sr)VO$_3$, which is also
considered as a bandwidth controlled system.\cite{Makino1998} Inset in Fig. 
\ref{mass} shows spectra of $1/\tau (\omega )$ for our (Ca,Sr)RuO$_3$ films.
The frequency dependence is almost linear up to 1 eV and the slopes are
nearly the same for all samples. This linear frequency dependence has been
observed in many other correlated systems,\cite{Makino1998,Schlesinger1990}
however, there is still no concrete explanation for this behavior.

\begin{figure}[htbp]
\epsfig{file=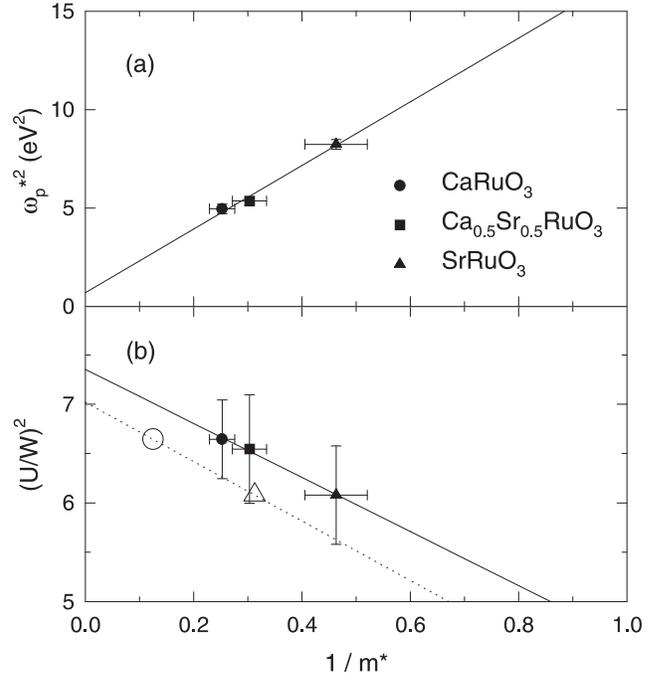,width=3.3in,clip=}
\vspace{2mm}
\caption{(a) Spectral weight of the QP peak, $\omega _p^{*2}$, versus $%
1/m^{*}$. (b) $(U/W)^2$\ versus $1/m^{*}$. Solid symbols represent the estimated values
of $1/m^{*}$\ at $\omega =$ 0.23 eV. For the open symbols, i.e. $%
\bigcirc $\ (CaRuO$_3$) and $\bigtriangleup $\ (SrRuO$_3$), the dc values
from specifit heat measurements are used.}
\label{fit}
\end{figure}

Within the GBR picture for the Mott transition, the spectral weight of the
QP peak, $\omega _p^{*2}$, is an order parameter and should be proportional
to $1/m^{*}$. Using a relation such that 
\begin{equation}
\omega _p^{*2}\simeq 8\int_0^{1\ {\rm eV}}\sigma (\omega )d\omega ~,
\label{order}
\end{equation}
$\omega _p^{*2}$\ values for the (Ca,Sr)RuO$_3$ films could be estimated. As
shown in Fig. \ref{fit}(a), the measured values of $\omega _p^{*2}$ are
quite linear to $1/m^{*}(\omega =$ 0.23 eV$)$ with a very small y-axis
intercept, demonstrating that most of free carriers in the (Ca,Sr)RuO$_3$
films are correlated QP's. Fig. \ref{fit}(b) shows the dependence of $%
(U/W)^2 $ on $1/m^{*}(\omega =$ 0.23 eV$)$. From Eq. (\ref{GBRmass}), $%
(U/W)^2$ should be proportional to $(-1/m^{*})$, and the y-axis intercept
should provide information on $(U/W)_c$. With $m^{*}(\omega =$ 0.23 eV$)$, $%
(U/W)_c$ was estimated to be about 2.7. When we use dc values of $m^{*}$,
which were shown with the open circle and triangle, $(U/W)_c$ was estimated
to be slightly lower by 0.05. It should be noted that the linear behaviors
in Figs. \ref{fit}(a) and \ref{fit}(b) clearly demonstrate a
self-consistency of our analysis based on the correlation picture.

Note that the measured value of $(U/W)_c$ $\simeq $ 2.7 for (Ca,Sr)RuO$_3$
is much larger than that of quantum Monte Carlo (QMC) simulation result,
i.e. 1.5, for the half-filled single band Hubbard model.\cite{Georges1996}
Reported general trends are that the orbital degeneracy increases $(U/W)_c$,
and that the electron filling away from the half-filling decreases $(U/W)_c$.%
\cite{Lu1994Gunnarsson1996} From our QMC simulations for triply degenerate
cases, it was found that $(U/W)_c$ $\simeq $ 2.3 for half-filling, 2.0 for
2/6- and 4/6-fillings, and 1.9 for 1/6- and 5/6-fillings.\cite{analytic}
Even after the orbital degeneracy and the filling factor are considered
properly, the QMC result seems to be smaller than our experimental value.
Some portion of this discrepancy might come from the uncertainty in $W$,
experimentally deduced from the half width of $\sim $ 3 eV optical
absorption band. A similar problem existed in the metallic $3d$ (Ca,Sr)VO$_3$
cases, even though it was not clearly stated.\cite{Rozenberg1996}

The QMC result for 4/6-filling is shown in the inset of Fig. \ref{cond}(b).
It is clearly seen that the predicted width of the QP peak in $\sigma
_1(\omega )$\ is much narrower than the measured value.\cite{DMFT} More
theoretical considerations are required to explain our results for QP peak
and $(U/W)_c$. Also, in (Ca,Sr)RuO$_3$, there might be other kind of
interactions, including $p$-$d$ hybridization, impurity scattering, and
magnetic fluctuation due to double-exchange interaction, etc. The interplay of
such interactions with Mott-Hubbard physics might provide us some more
insights to understand the discrepancy of $(U/W)_c$, and even to explain why
correlation effects appear to be so strong in $4d$ electrons of
ruthenates.

In summary, we found that (Ca,Sr)RuO$_3$ is a $4d$ metal compound which has
strong electron correlation effects. Its optical spectra display systematic
changes in the spectral characteristics predicted for a metallic
Mott-Hubbard system.

This work was supported by the Ministry of Education through BSRI-97-2416 and by
the Korea Science and Engineering Foundation (KOSEF) through the RCDAMP of Pusan
National University. One of us (JB) was supported by the KOSEF through
the Postdoctorial Fellowship Program.


\begin{references}
\bibitem{m*}  $m^{*}$ is normalized quantity with respect to band mass for
an uncorrelated system.
\bibitem{Wp}  For the calculation, plasma frequency $\omega _p$ = 3.9 eV is
used.
\bibitem{analytic}  On the other hand, 3.65 is obtained by Lu\cite{Lu1994Gunnarsson1996} from
the analytical calculations for 4/6-filled triply degenerate case, within the
Gutzwiller approximation.
\bibitem{DMFT}  For the calculation, we used the dynamic mean field theory
with $T$ = 1/12 (in the unit of $W/2$ = 1).
\end{references}
\end{document}